\documentclass[10pt, paper=a4, UKenglish]{article}
\usepackage{graphicx}
\def\Title#1{\begin{center} {\Large #1 } \end{center}}
\def\Author#1{\begin{center}{ \sc #1} \end{center}}
\def\Address#1{\begin{center}{ \it #1} \end{center}}

\newcommand\pubblock{\rightline{\begin{tabular}{l} Proceedings of the CTD/WIT 2019\\ \pubnumber\\
         \pubdate  \end{tabular}}}

\newenvironment{Abstract}{\begin{quotation} \begin{center} 
             \large ABSTRACT \end{center}\bigskip 
      \begin{center}\begin{large}}{\end{large}\end{center} \end{quotation}}

\newenvironment{Presented}{\begin{quotation} \begin{center} 
             PRESENTED AT\end{center}\bigskip 
      \begin{center}\begin{large}}{\end{large}\end{center} \end{quotation}}

\def\beq{\begin{equation}}
\def\eeq#1{\label{#1}\end{equation}}
\def\eeqn{\end{equation}}

\def\beqa{\begin{eqnarray}}
\def\eeqa#1{\label{#1}\end{eqnarray}}
\def\eeqan{\end{eqnarray}}

\let\bar=\overbar

\def\Dslash{\not{\hbox{\kern-4pt $D$}}}
\def\dslash{\not{\hbox{\kern-2pt $\del$}}}

\def\msb{{\bar{\ssstyle M \kern -1pt S}}}

\usepackage{color}
\usepackage{lineno}
\usepackage{subfig}
\usepackage{hyperref}

\newcommand\pubnumber{PROC-CTD19-102}

\newcommand\pubdate{\today}

\def\affiliation{
Institute for Information Processing Technologies \\
Karlsruhe Institute of Technology, Germany \\}

\def\affiliationM{
Max-Planck-Institute for Physics, Munich, Germany\\}

\newcommand{\conference}{Connecting the Dots and Workshop on Intelligent Trackers (CTD/WIT 2019)\\
Instituto de F\'isica Corpuscular (IFIC), Valencia, Spain\\ 
April 2-5, 2019}

\usepackage{fancyhdr}
\definecolor{mygrey}{RGB}{105,105,105}
\begin{document}


\large
\begin{titlepage}
\pubblock

\vfill
\Title{Low Latency Neural Networks using Heterogenous Resources on FPGA for the Belle II Trigger}
\vfill

\Author{Steffen Baehr, Sara McCarney, Felix Meggendorfer, Julian Poehler, Sebastian Skambraks, Kai Unger, Juergen Becker,Christian Kiesling}
\Address{\affiliation}

\Address{\affiliationM}
\vfill

\begin{Abstract}
One of the major components of the Belle II trigger system is the neural network trigger. Its task is to estimate the z-Vertex particle tracks observed in the experiments drift chamber. The trigger is implemented on FPGAs to ensure flexibility during operation and leverage their IO capabilities. Meanwhile the implementation has to estimate the vertex in a few hundred nanoseconds to fulfil the requirements of the experiment. A first version of that trigger was operational during the first collisions. While it was able to estimate the vertex, it had some drawbacks regarding the possible throughput and timing closure. These are the focus of this work, which modifies the original design to allow two networks running in parallel and less routing congestion. We conducted a rescheduling of multiply and accumulate which are the basic operations in such networks. While the original design tried to parallelize as much as possible, the rescheduling tries to reduce the number of parallel data transmission by reusing processing modules. This way resource consumption was reduced by 40\% for DSPs. To further increase the throughput by operating an additional network in parallel, we investigated the balanced use of SRAM-LUTs and DSPs for multiply and accumulate operations. With the found balancing ratio the trigger is able to operate two neural networks in parallel on the targeted FPGA within the required latency.
\end{Abstract}

\vfill

\begin{Presented}
\conference
\end{Presented}
\vfill
\end{titlepage}
\def\thefootnote{\fnsymbol{footnote}}
\setcounter{footnote}{0}
%

\normalsize 


\section{Introduction}
\label{intro}

The Belle II particle accelerator experiment located at Tsukuba, Japan is targeting a collision rate that is 40 times higher than its predecessor Belle~\cite{BelleII}.  It is an asymmetric experiment with separate two beams of positrons and electrons. The increase of the collision rate is necessary in order to find new physics beyond the standard model. This, however, has a significant impact on readout system of the detector’s as the number of collisions increases the amount of data to be stored. The resulting data rates easily reach the point at which full readout of the detector cannot be performed in a cost-efficient way as expensive data transmission infrastructure has to be developed and installed. Relief is typically provided by using online processing system close to the detector readout that compress, reduce and limit the trigger data of events. Reduction and trigger approaches are based on the fact that the origin of most observed particle tracks is not from a collision but rather a result of background. When data belonging to background events can be distinguished from physics events, readout can be limited to send physics events without losing valuable data. Performing this task as early is possible in the data readout scheme of a detector, results in the highest potential for reduction of data rates to be stored. However, they require close integration with the respective detector and consideration of its distinct behaviour during operation. In the case of Belle II this dependency requires the development of new trigger solution instead of completely reusing the approaches from the predecessor.
\newline
Due to the amount of particle tracks to be observed it is foreseen that the resulting data rates for Belle II are going to be higher than the available transmission bandwidth to the offline computing facilities. As a result, only part of the whole detector data can be sent. This is a common problem in modern high energy particle physics experiments and is typically solved by using a triggered data readout. Such a scheme uses a trigger system that is deciding online during the experiment at which point in time the complete detector data is sent out or not, thus managing the readout frequency of an experiment. Here, FPGAs are used to achieve the necessary bandwidth and latency requirements

\begin{figure}[h]
	\centering
	\includegraphics[width=0.5\linewidth]{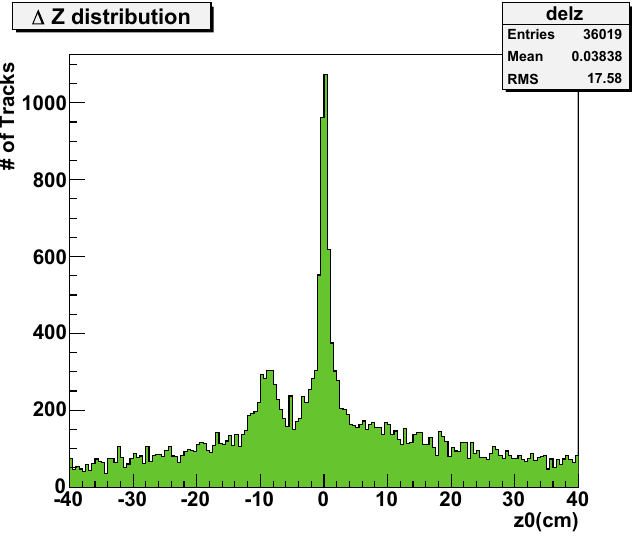}\caption{$z$-Distribution from the Belle experiment showing peaks outside of $z=0$.}
	
	\label{fig:zVertex}
	
\end{figure}

Already in the predecessor Belle, it was observed that a high amount of data that was read out belonged to particles with an origin outside of the interaction point (IP) at $z = 0$. The distribution of events and their reconstructed origin along the $z$-Axis is shown in figure~\ref{fig:zVertex}. A significant part of the available data rate was thus used for particles that do not contribute to new physics. With the redesign of the trigger system in Belle II the goal to suppress these particles was defined.
\newline
The observable peaks outside of $z = 0$ motivate the development of a neural z-Vertex trigger that is capable estimating the parameter using neural networks. It is an FPGA based trigger system that uses a multilayer perceptron network (MLP) that was trained to estimate the z-Vertex. To stay within the latency budget of the entire trigger system as well as match the throughput requirements the estimation has to be made available within an overall latency of $5 \mu s$ while achieving a throughput frequency of $31.75$ MHz~\cite{L1Trg}.

This paper is organized in the following way. Section~\ref{sec:belle} provides context about the Belle II detector and its L1 trigger system that are important to understand the system described in here.  In Section~\ref{sec:soa} discusses related work of online track trigger concepts and design flows for neural networks on FPGAs. The functional description and architecture of the trigger is presented in section~\ref{architecture}, followed up by the methods and techniques to implement low latency and resource efficient implementation of neural networks in section~\ref{implementation}. An evaluation of the system's characteristics in terms of resource consumption is presented in section~\ref{results}. Section~\ref{conclusion} concludes the paper with a discussion and outlook.

\section{Belle II Detector}
\label{sec:belle}

The essential part of the observation of collisions is an exact reconstruction of the subsequent decays. The experiment uses three detectors, which are primarily used for reconstruction of particles tracks. These are the pixeldetector (PXD), silicon vertex detector (SVD) and central drift chamber (CDC). The PXD and the SVD are completely new developments for the Belle II experiment, while the CDC is mostly reused from Belle. The PXD forms the innermost detector of the experiment and is placed directly around the IP. It provides a much higher precision in spatial measurements, but has a relatively long integration time due to the used DEPFET sensor technology. The SVD is enclosing the PXD and is in contrast it is a strip detector with less accuracy in spatial resolution but a shorter readout time and higher coverage of space. Both are enclosed by the CDC that is mainly used to determine a track's momentum and charge by estimating the track parameters. These three detectors are used to determine decay nodes and find tracks with low momentum. The tracking detectors are enclosed by particle identification detectors. The time of propagation detector is located directly after the CDC at the barrel while the particle identification detector is installed at the endcaps. These detectors are enclosed by the energy and cluster Detector, which is used for detection of photons and the identification of electrons. Finally, the kaon and muon Detector is used to identify $K^0_{L}$ and muons.
\newline
The CDC~\cite{CDC} is hereby the most important for this paper. It is constructed according to a layered model. Each layer consists of a number of wires stretched parallel to the $z$-Axis and surrounded by a gas mixture. This gas mixture consists of a so-called low-Z gas (50\% each of helium and ethane). Interactions of the passing particles with the gas mixture is producing charge carriers at the wires. These are then used to detect the passing of a particle. The wires are grouped in layers, in total 56 layers are forming the entire detector. Successive layers are arranged with an increasing distance to the IP. Additionally, the number of wires varies across layers. Layers that are placed further away contain more wires, to cover the correspondingly larger space to be covered for detection of particles.

\begin{figure}[h]
	\centering
	\includegraphics[scale=0.9]{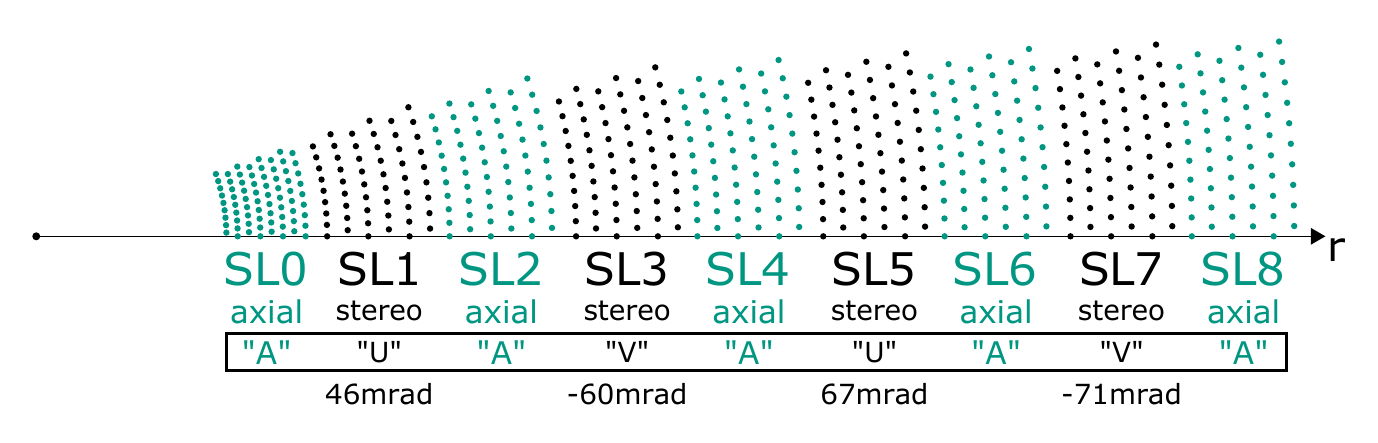}\caption{Alignment and configurations of the SuperLayers in the CDC.}
	
	\label{fig:cdc:struct}
	
\end{figure}

The layers are grouped together into nine SuperLayer (SL) that are enumerated from zero to eight. All wires have properties specific to their SL. They have an orientation, in which there are either aligned parallel to the $z$-axis, called axial. Or they have a stereo orientation in which they are aligned with an angle between $45.4$ to $74.0$ mrad to the $z$ axis depending on the SL. This angle is introduced to enable three-dimensional reconstruction of tracks. The entire arrangement is shown in figure~\ref{fig:cdc:struct}.

The trigger system of the CDC is the most important sub-trigger of Belle II. It has the longest processing pipeline and therefore requires the highest overall latency to generate the necessary signals. For the entire processing from frontend to global decision logic (GDL) including all communication $5 \mu s$ is fixed, while it has to generate trigger signals with a frequency of $30$ kHz. The NNT developed in this paper is one of its sub-systems. To understand the background and resulting requirements, the system architecture is outlined.

\begin{figure}[ht]
	\centering
	\includegraphics[scale=0.5]{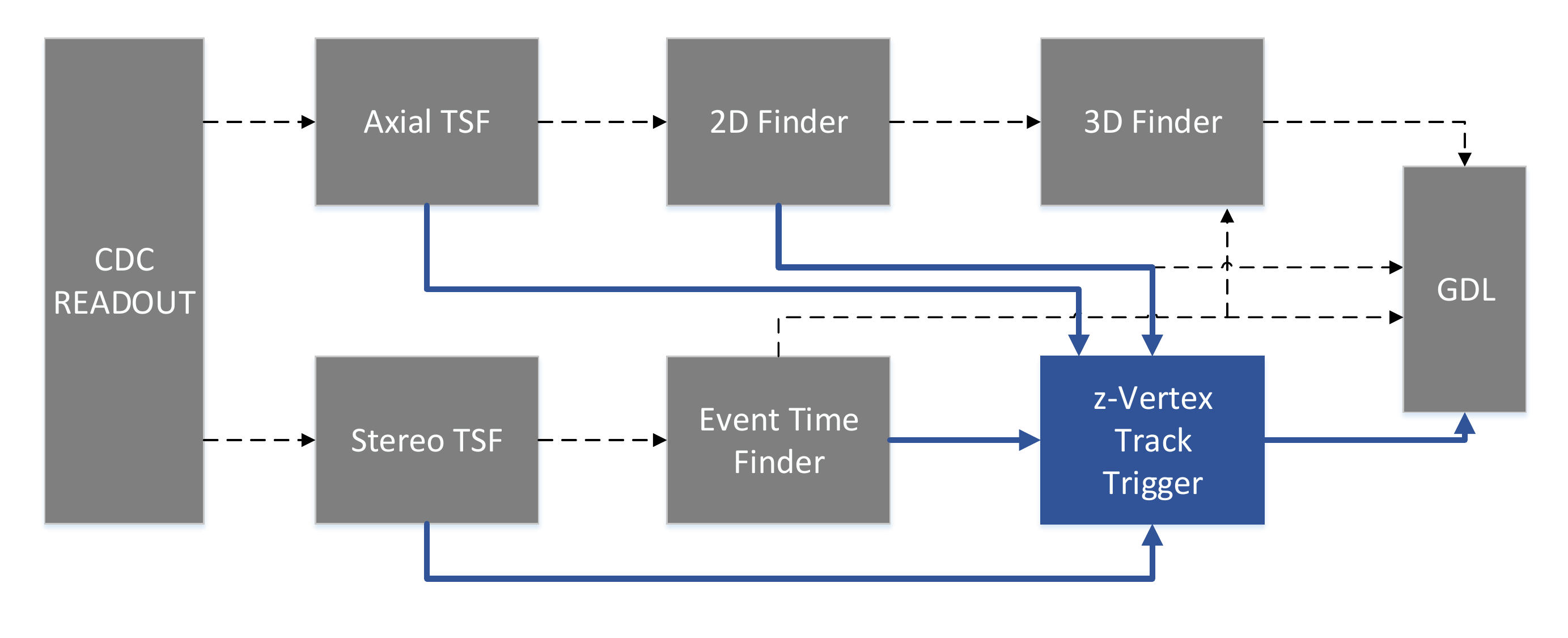}\caption{System architecture for the L1 Trigger of the CDC.}
	
	\label{fig:cdc_neuro_archi}
	
\end{figure}

Overall a multi-level and pipelined processing architecture is used. Data of individual SLs is continuously read out by the FrontEnd. It is then transferred to merger units, which concentrate data of the entire CDC. The concentrated data is then sent to the first stage of the trigger logic, the track segment finder (TSF)
The task of the TSF is to combine the individual wires into so called track segments (TS). These are generated for each SL separately in parallel by dedicated FPGAs. TS are additionally distinguished into axials and stereo according to the orientation of the wires within the processed SL. Axials are first processed by the 2D track finder, which tries to find a suitable 2D-tracks. At the same time using the CDC an event time by an event time finder. The outputs from these two modules together with the stereo TS are then sent to the 3D-Track Finder and the NNT. Both of these have the responsibility to send an estimation of the $z$-Vertex to the GDL. The architecture is presented in figure~\ref{fig:cdc_neuro_archi}.
 
\section{State of the Art}
\label{sec:soa}
With the usage of neural networks, the relationship based on data taken from the experiment or simulation can be learned without the necessity to create a precise algorithm. Another popular approach is the usage of a Hough transformation~\cite{CMS} or linear regression for track finding. These approaches extrapolate tracks by using single detector hits and fitting a track around them. They are viable options for readout triggering, achieving accurate results in the past. However, since the background is at the moment not completely understood, simulations showed that neural networks are projected to be more robust for future deployment. Additionally, adjusting these algorithms to the changing behaviour of the experiment is not an easy task, since they have to be redesigned. Neural networks on the other hand can be simply retrained using newly acquired data. Since neural networks offer the described advantages compared to traditional approaches, they were already used in past experiments for example the H1-Level 2 Trigger of the HERA experiment~\cite{Hera1}. However, that trigger was implemented on dedicated processors and had a rather large time budget of $200 \mu s$ compared to the requirements of the Belle II L1 trigger at $5 \mu s$.

Employing neural networks on FPGAs gained significant traction over the last years. FINN~\cite{FINN} showed how an abstract network description can be transferred efficiently onto FPGAs. It is nowadays used as reference for further improvements. However, these frameworks are not meant to be used for high energy physics use cases as they don’t provide the necessary interfaces are not targeted at their specific low latency requirements. They rather focus on predefined architecture that cover a broad range of applications. The lack of a suitable design flow for this domain was recognized with HLS4ML which is a framework that can generate a neural network architecture~\cite{Durate} based on a physics use case. However, it is more focused on the investigation of feasibility of neural networks for trigger systems in general, while the presented work represents a complete system that is currently in use and integrated into the data flow of the experiment.

\section{Neural z-Vertex Trigger}

Algorithmically the employed approach for estimating the $z$-Vertex consists of two main stages~\cite{Skambraks}. At first a preprocessing stage transforms the detector’s data into a more viable representation. Using this the neural network is estimating of the $z$-Vertex. Often the preprocessing stage can be omitted and compensated by using a more complex network. Since both the latency and resource budget is tightly constrained, a preprocessing stage that makes use of the geometrical information of the CDC is used reducing the number of necessary networks and neurons. Reasonable accuracy for the estimation can already be achieved using just one network. However, the final implementation uses a total of five networks that can compensate missing wires in the stereo SLs. In case one of the four stereo layers does not have a matching TS, a specialized network is loaded during runtime.

\begin{figure}[h]
	\centering
	\includegraphics[scale=0.7]{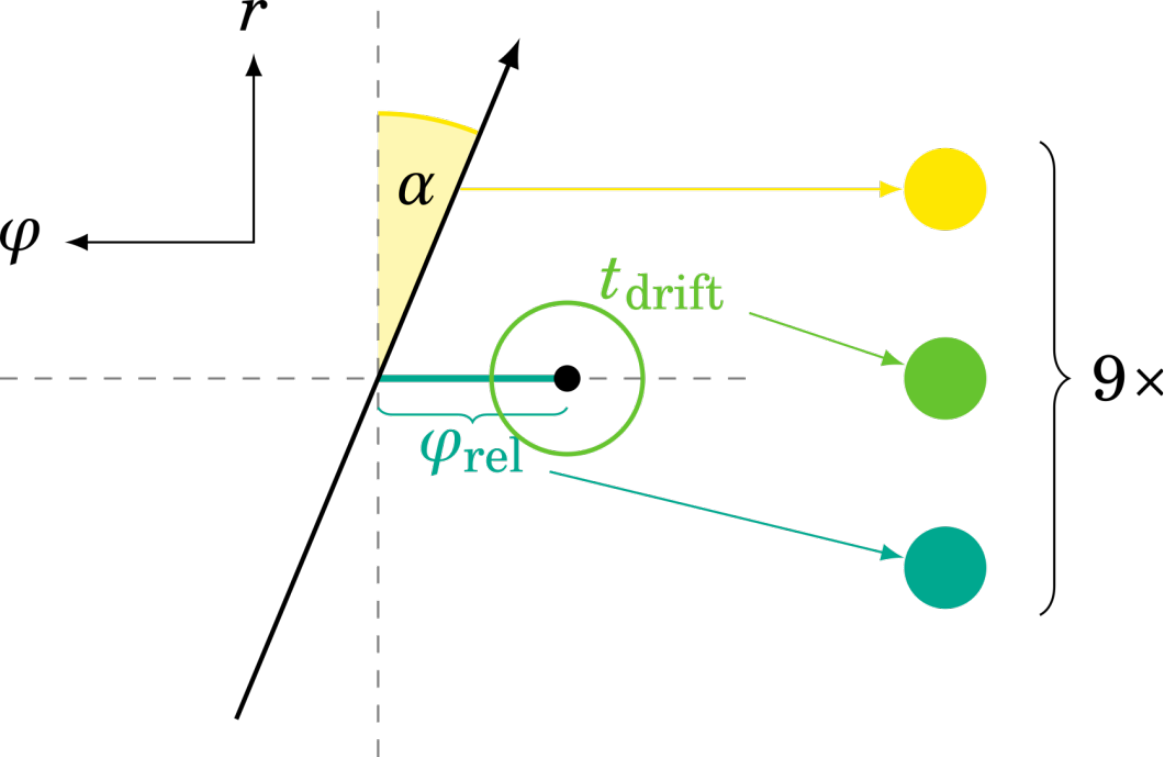}
	\caption{Graphical representation of the inputs to the neural network relative to a wire of the CDC~\cite{Sara}.}
	\label{fig:preprocessing}
	
\end{figure}

The preprocessing calculates three separate input variables for each of the SLs of the CDC. This results in 27 inputs in total for the network.  The three variables are the crossing angle $\alpha$, which is between the estimated 2D particle track and an active reference wire of the CDC, the wire’s position relative to the estimated 2D track $\phi_{rel}$ and the drift time that describes the time between the particle passing through the space and a hit being registered at a wire. The geometrical representation of the inputs is shown in figure~\ref{fig:preprocessing}.

For the network itself an MLP with one input, hidden and output layer is used. Each node in the hidden and output layer represents a neuron that computes a weighted sum over its input values. From a processing point of view, the MLP mainly consists of a set of multiply and accumulate operations (MAC). For each neuron's output a predefined activation function is applied, for which we use the hyperbolic tangent. In our setup all input and output values of the MLP are scaled to values within the range [$-1$,$ 1$].
\newline
Training of the MLP showed that using 81 neurons for the hidden layer and two neurons for the output layer is sufficient to achieve the required results for estimating the z-vertex. 
Our network training is performed using the iRPROP algorithm~\cite{iRPROP} which is an improved variant of the RPRO backpropagation algorithm. This algorithm has been demonstrated to be more effective than the classical backpropagation, because the magnitude of the weight update is independent of the magnitude of the derivative of the cost functions and only dependent on the dynamics of the past weight updates. The effect is a faster convergence to a minimum of the cost function with the same minima found compared to the classical backpropagation.

\label{architecture}
\begin{figure}[h]
	\centering
	\includegraphics[scale=0.7]{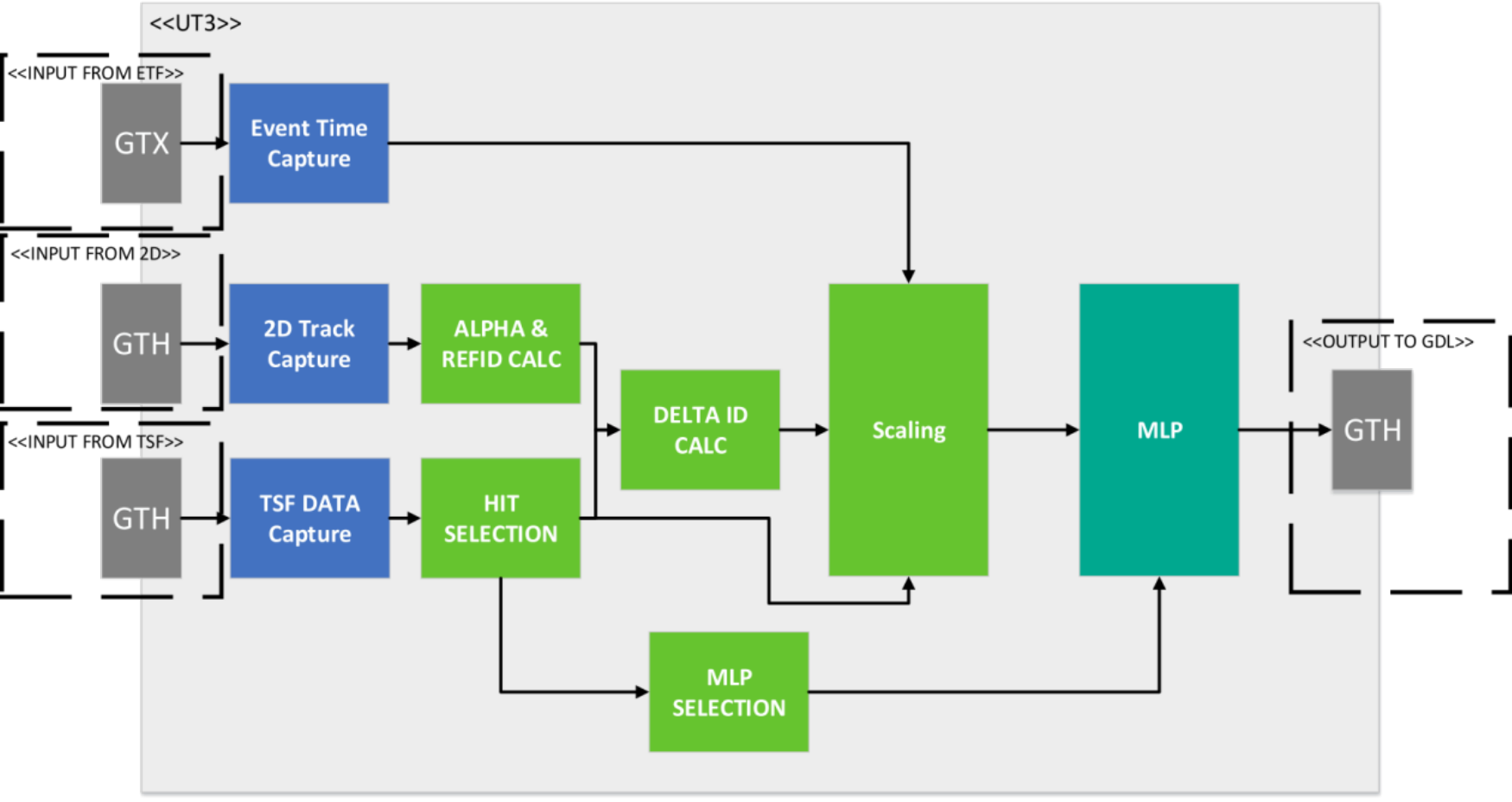}
	\caption{Architecture of the implementation on the FPGA. IO resources are colored grey, modules responsible for data transmission protocols in blue, while preprocessing modules are light green and the MLP is in dark green.}
	\label{fig:archi}
	
\end{figure}

The architecture of the FPGA implementation is shown in figure~\ref{fig:archi}. It is divided into three stages. The input handling that receives the data from the different sources within the CDC trigger system. The preprocessing which is represented by the different processing modules used within the design. These additionally show the data flow related dependencies in processing. Modules without a data dependence are operating in parallel and data is synchronized to compensate for different delays at the respective stages. The main core of this paper is the MLP, which is the last part of processing.

\section{Implementation Techniques for Neural Networks}
\label{implementation}
The core of a neural network is the artificial neuron. Algorithmically it basically consists of a set of MACs. Here the number of operations per neuron is equal to the number of inputs. An optional bias may be used but it does not require additional multiplications since it can be added as a constant. At this stage there are already many degrees of freedom for the implementation on an FPGA. The optimal implementation variant depends on the characteristics of the network. Parameters like the bit widths used for weights and input values are important here. The same applies to whether weights are static or shall be reloaded during operation. Low-latency and resource demanding implementation is addressed here with two techniques, usage of heterogeneous resources and scheduling of operations based on time multiplexing.

\subsection{Heterogeneous Implementation}

Modern FPGAs provide dedicated DSP slices that are particularly suitable for a realization of MACs in which multiple different weights can be loaded during runtime. These processing elements already have an integrated optimized multiplier for variable inputs on the chip.  In addition, a slice already contains an accumulator so that both processing steps can be implemented in the same component. The optimized multiplier allows a DSP to achieve higher throughput and lower latency. At the same time, multiplications with variable weights can only be realized more inefficiently in terms of resources in SRAM.

However, if the weights are constant, a realization in SRAM might be the better option. Instead of providing resources for general multiplication, only this constant is realized. In addition, SRAM based implementation can be routed easier and more thoroughly logic optimized due to their large-area availability and finer granularity. An implementation of a neuron is also strongly dependent on the selected bit widths. A DSP slice has fixed bit widths for both input ports. On a Virtex-6, these are limited to 18 and 21 bits for the two input ports. As soon as the bit widths are exceeded, additional DSPs are required. In addition to the additional resources required, this also increases latency due to the needed communication combining partial results of multiple DSPs. 

A general disadvantage of an implementation with DSPs is that their limited availability on an FPGA. Additionally, they have a fixed location arranged in columns on modern architectures. These columns may be far away from the data sources, routing the signals to these might exceed the targeted timing. When DSP slices are reused across several data sources at different clock cycles, routing resources might get overcrowded and congestion occurs which limits the achievable frequency. Due to this disadvantage we implemented the option to choose between the implementation of MACs in SRAM-LUT and DSPs. This allows to balance out the demand across all available processing resources. From an implementation point of view, this is realized by defining utilization ratios. For example, if the ratio can be set to 40\% resulting in that fraction of MACs being executed on SRAM-LUT instead of DSPs.

\subsection{Time Multiplexing of MACs and Neurons}

A good trade-off between latency, resources and the timing closure can be achieved by employing time multiplexing for the execution of MACs of a neuron. Instead of allocation exactly one DSP or SRAM-LUT based module to each input of a neuron, it can be designed to only process a subset of inputs for every clock cycle. Operations that were not processed immediately are then postponed to later clock cycles.

\begin{figure}[ht]
	\centering
	\includegraphics[width=0.8\linewidth]{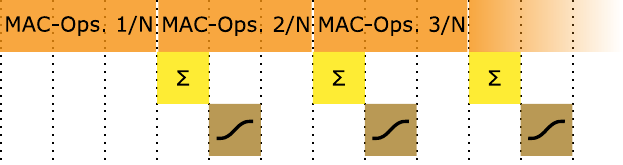}\caption{Schedule of a time multiplexed DSP. MAC operations of the different neurons are executed on one DSP unit at different time intervals. Both the adding and activation function can be interleaved with the execution of the next neuron.}
	
	\label{fig:muxNeuron}
\end{figure}

Using time multiplexing will increase the latency for processing an entire neural network, but offers the opportunity to significantly reduce required resources. The effects on the routing and the size of the adder tree are also interesting here. Additionally, in case of realization on DSPs when several inputs of the same neuron are processed on the same DSP, the internal accumulator can be used to already form partial sums. Then the following adder tree only has to add the partial sums instead of all. This further reduces resources demand in terms of routing and stages within the adder tree. This is shown in figure~\ref{fig:muxNeuron}, in which dashed lines indicate the orders of clock cycles. Here, MACs of one neuron are processed in three successive clock cycles. In addition to the increase in latency, this approach also requires additional registers for storing of input values and possibly weights if these are not constant, as well as multiplexers. Furthermore, a controller for directing the multiplexers across clock cycles is necessary.

Time multiplexed operation is not restricted to MACs within a neuron, neurons themselves can also be processed that way. In this case, neurons of the same network are processed on the same resources at different clock cycles. The advantages and disadvantages are the same as for the inputs. Both variants are shown in a schedule in figure~\ref{fig:muxNeuron}. In this schedule, the operations are multiplexed in time by processing all input values of a neuron in three consecutive clock cycles. Then another neuron is processed. At the same time the adder tree and the activation function can be executed overlapping to the multiplications of another neuron.

\begin{figure}[h]
	\centering
	\includegraphics[scale=0.8]{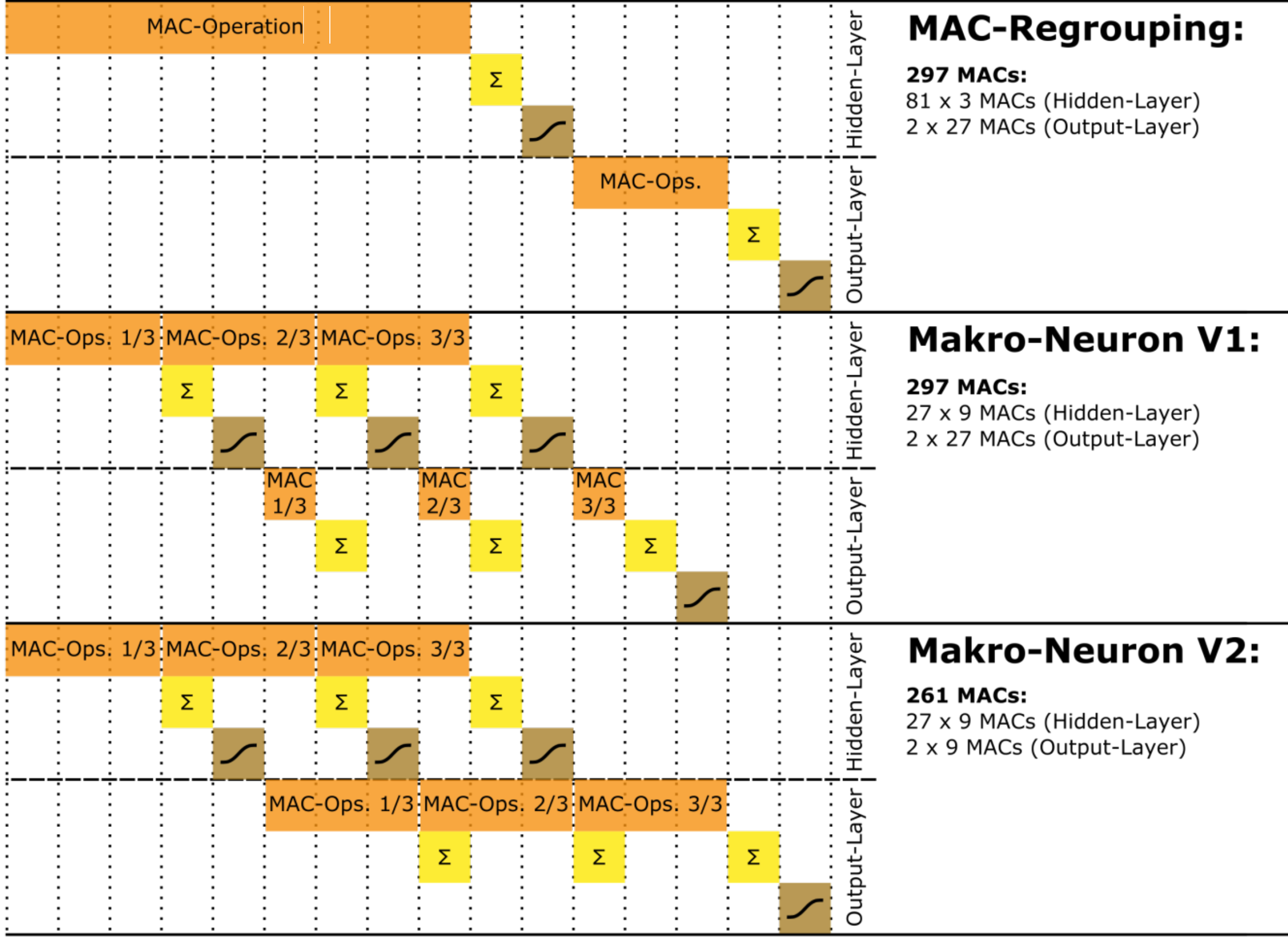}
	\caption{Impact of the pipelining and scheduling options on resources and latency.}
	\label{fig:pipe}
	
\end{figure}

\section{Results}
\label{results}
Results of the used approaches can be presented in terms of resource consumption and performance. Performance is meanwhile both latency and throughput. Those are however fixed constraints by the experiment and are mandatory to be fulfilled. The optimizations are thus rather targeting resource consumption. Figures of merit are based on the universal trigger board 4, which hosts a Virtex Ultrascale XCVU080~\cite{Virtex}. This trigger board will be the platform to be used in future stages of the experiment and provides the environment for the optimizations presented here.
\newline
The impact of the presented techniques that are aiming at rescheduling the execution of the neural network, are shown in figure~\ref{fig:pipe}. Since MACs can be implemented in both SRAM-LUT and DSPs, resources will be represented by MAC units, which are rather required parallel modules then the low-level FPGA resource. Using time multiplexing of both MACs and neurons can bring down the resource consumption for DSPs to required 297 MAC-Units. When execution of different layers of the network is interleaved and pipelined the overall latency can be reduced. As these additional clock cycles do not have a benefit, since execution just has to stay within the fixed latency, execution of neurons from the output layer can be stretched across additional clock cycles. While this negates the gained advantages of interleaving in terms of latency, resource consumption is further decrease as less MAC-Units need to be used.
\newline
A comparison between an implementation using both DSPs and SRAM-LUT for MACs and only DSPs is shown in table~\ref{tab:resources}. The ratios for which the MACs are divided across both types of resources are 40\% implemented in SRAM-LUT and 60\% in DSPs. In the DSP only implementation only a small part of the SRAM-LUTs are actually in use, only about 17\%. Those mainly in use for Adder-trees combining partial sums of separate MACs, the activation function and multiplexers. For our chosen network topology about 68 \% of DSPs are used, however these are more difficult to route. By using the heterogeneous approach, resource demand is more evenly distributed across all types as about 30\% of the SRAM-LUTs are now utilized while DSPs are reduced to 44\%. This typically decreases the maximum frequency of a design; however, it still fits the 127MHz clock used for the design.

\begin{table}[!htb]
  \begin{center}
    \begin{tabular}{l|ccc}
      \hline
      \hline
      Variant &  LUT & FF &  DSP \\
      \hline
      DSP only &  17\%  &  2\% & 68\% \\
      Heterogeneous &  30\%  &  3\% & 44\%  \\
      \hline
      \hline
    \end{tabular}
    \caption{Resource utilization for heterogeneous and DSP only implementation of the processing logic.}
    \label{tab:resources}
  \end{center}
\end{table}


\section{Conclusions}
\label{conclusion}
The focus of this paper is the optimization of the neural network implementation on FPGAs used in the neural $z$-Vertex trigger that is used in the Belle II experiment. The optimizations can be categorized into scheduling and usage of heterogeneous resources. With different scheduling approaches the resources necessary to perform the neural networks can be reduced, while both latency and throughput requirements are fulfilled. Additionally, this has a positive effect on the FPGA-specific implementation, as execution of neurons in separate clock cycles can reduce the routing demand, as input values for a neuron can be reused instead of routed in parallel to multiple MAC units. Additionally, using both SRAM-LUTs and DSPs for MACs can more evenly distribute resource demand across the available types present on FPGAs. As the neural $z$-Vertex trigger is already in operation without these techniques, they are especially of interest for an upgrade, in which multiple networks are implemented and operated in parallel on an FPGA, in order to increase the number of tracks that can be processed.




\begin{thebibliography}{99}



\bibitem{BelleII}
T. Abe et al. , ”Belle II Technical Design Report" , ArXiv Nov. 2010

\bibitem{L1Trg}
Y. Iwasaki et al., ”Level 1 trigger system for the Belle II experiment”, IEEE Trans. Nucl. Sci. 58 (2011) 1807

\bibitem{CDC}
N, Taniguchi et al., "Central Drift Chamber for Belle-II. Journal of Instrumentation", 12(06):C06014, 2017.

\bibitem{CMS}
C. Amstutz et al., "An FPGA-based track finder for the L1 trigger of the CMS experiment at the high luminosity LHC", 2016 IEEE-NPSS RT

\bibitem{Hera1}
A. Gruber et al, ”A neural network architecture for the second level trigger in the H1-experiment at the electron proton collider HERA”,
arXiv:1406.3319 [physics.ins-det]

\bibitem{FINN}
Y. Umuroglu, ”FINN: A Framework for fast scalable binarized Neural
Network inference”, arXiv:1612.07119

\bibitem{Durate}
J. Duarte et al., ”Fast inference of deep neural networks in FPGAs for particle
physics”, arXiv:1804.06913

\bibitem{Skambraks}
S. Skambraks et al., ”A z -Vertex Trigger for Belle II”, [arXiv:1406.3319]


\bibitem{Sara}
S. Pohl, "Track vertex reconstruction with neural networks at the first level trigger of Belle II", Disseration Ludwig-Maximilians-University Munich

\bibitem{iRPROP}
C. Igel et al., “Improving the Rprop Learning Algorithm”, Proceedings of the Second Int. Symposium on Neural Computation, NC 2000

\bibitem{Virtex}
XILINX Inc., "UltraScale FPGA Product Tables and Product Selection Guide", 2018.


\end{thebibliography}
\end{document}